%% file: copula_forecast.tex
\documentclass[12pt,english]{article}
\usepackage{avant}
\usepackage[T1]{fontenc}
\usepackage[utf8]{inputenc}
\usepackage{geometry}
\usepackage{babel}
\usepackage{amsmath}
\usepackage[authoryear]{natbib}
\usepackage{amssymb}
\usepackage{stmaryrd}
\usepackage{algorithm}
\usepackage{algcompatible}
\usepackage[]{algpseudocode}
\usepackage{graphicx}
\usepackage{rotating} 
\usepackage{float}
\usepackage{booktabs} 
\usepackage{subcaption}
\usepackage{multirow}
\usepackage{multicol}
\usepackage{setspace}
\usepackage{threeparttable}
\usepackage[unicode=true,
bookmarks=false,
breaklinks=false,pdfborder={0 0 1},backref=false,colorlinks=false]
{hyperref}
\usepackage{breakurl}

\makeatletter

\usepackage{babel}
\usepackage{array}
\newcolumntype{L}[1]{>{\raggedright\let\newline\\\arraybackslash\hspace{0pt}}m{#1}}
\newcolumntype{C}[1]{>{\centering\let\newline\\\arraybackslash\hspace{0pt}}m{#1}}
\newcolumntype{R}[1]{>{\raggedleft\let\newline\\\arraybackslash\hspace{0pt}}m{#1}}

\floatstyle{ruled}

\usepackage{amsthm}\usepackage{bm}\usepackage{color}
\usepackage{multirow}\usepackage{enumerate}\usepackage{caption}\usepackage{subcaption}

\floatstyle{ruled}

\topmargin by -\headheight \advance
\topmargin by -\headsep     

\evensidemargin
\oddsidemargin
\@ifundefined{definecolor}{color}{}
\RequirePackage[displaymath]{lineno}

\@ifundefined{definecolor}{color}{}
\usepackage{amsfonts}\usepackage{latexsym}

\usepackage{mathrsfs}\usepackage{amsthm}\usepackage{latexsym}\usepackage{booktabs}\@ifundefined{definecolor}{color}{}

\usepackage{times}\@ifundefined{definecolor}{color}{}

\newtheorem{definition}{{\bf Definition}}
\newtheorem{theorem}{{\bf Theorem}}

\usepackage{babel}

\makeatother

\begin{document}
\input{setting}
\title{Forecasting Realized Volatility Matrix With Copula-Based Models}
		\author{Wenjing Wang and Minjing Tao\footnote{To whom correspondence should be addressed: Minjing Tao (tao@stat.fsu.edu).}
			\\ Department of Statistics, Florida State University}

	\maketitle
\begin{abstract}
	Multivariate volatility modeling and forecasting are crucial in financial economics. This paper develops a copula-based approach to model and forecast realized volatility matrices. The proposed copula-based time series models can capture the hidden dependence structure of realized volatility matrices. Also, this approach can automatically guarantee the positive definiteness of the forecasts through either Cholesky decomposition or matrix logarithm transformation.  In this paper we consider both multivariate and bivariate copulas; the types of copulas include Student's $t$, Clayton and Gumbel copulas. 
In an empirical application, we find that for one-day ahead volatility matrix forecasting, these copula-based models can achieve significant performance both in terms of statistical precision as well as creating economically mean-variance efficient portfolio. Among the copulas we considered, the multivariate-$t$ copula performs better in statistical precision, while bivariate-$t$ copula has  better economical performance.

	\vspace{5pt}
	\noindent \textbf{Key Words:} realized volatility matrix; copulas; time series forecasting.
\end{abstract}
\newpage{}

\section{Introduction}
Volatility estimation and forecast for financial market have significant importance in the fields such as portfolio allocation, risk management and asset pricing, etc. With the high-frequency data available, different approaches of nonparametric estimations of the volatility become very popular. The estimators of univariate integrated volatility include two-scale \citep{zhang2005tale} and multi-scale estimators \citep{zhang2006efficient}, realized kernel volatility \citep{barndorff2008designing} and   pre-averaging approach \citep{jacod2009microstructure, christensen2010pre}. Also, \cite{barndorff2004econometric} implemented the estimation of volatility matrix for the multivariate case. Other volatility matrix estimators, such as realized co-range \citep{bannouh2009range}, realized kernel volatility matrix \citep{barndorff2011multivariate}, multi-scale realized covariance \citep{zhang2011estimating} are also becoming increasingly popular.

With the estimation of volatility available, different time series models can be applied to the series of realized volatilities. For the univariate case, the fractionally integrated ARMA (ARFIMA) \citep{andersen2003modeling} and the heterogeneous autoregressive (HAR) model \citep{corsi2009simple} perform very well in empirical applications, both of which can capture the long-memory dependence in realized volatility while retain parsimony. For the multivariate case, problem rises of how to guarantee positive-definite forecasts of a realized volatility matrix. The multivariate modeling approach includes the Wishart Autoregressive (WAR) \citep{gourieroux2009wishart} and Conditional Autoregressive Wishart (CAW) \citep{golosnoy2012conditional}. Vector ARFIMA (VARFIMA) \citep{chiriac2011modelling} is another approach which is based on the Cholesky decomposition of the realized volatility matrix. \cite{bauer2011forecasting} transformed the realized covariance matrix using matrix logarithm function and then modeled the dynamics of log-volatility matrix with a latent factor model.

Both Cholesky decomposition and matrix logarithm transformation are commonly used tools to ensure the positive definiteness of the realized volatility matrix. However, the existing models based on these methods overlook the nonlinear correlation and dependence structure such as asymmetry and tail dependence between the elements, which are partly caused by the nature of the transformation. For example, each element of Cholesky factors depends in a nonlinear way on the corresponding realized volatility and all Cholesky elements from previous row. Among the models to capture dependence structure, copula is one of the popular methods that can model the dependence characteristics of nonlinear time series.  \cite{ibragimov2008copulas} showed that Clayton copula-based time series models can exhibit long memory properties. 

Therefore, copula is potentially a good method to model the realized volatility. \cite{sokolinskiy2011forecasting} proposed an approach based on bivariate copula to model the volatility in the univariate case. See also \citet{simard2015forecasting}. For the multivariate case, \cite{brechmann2016multivariate} proposed a dynamic framework for modeling and forecasting realized covariance matrices using vine copulas. This vine-copula approach is still based on bivariate copulas to connect residuals of the univariate models for each element of the Cholesky factors. In this paper, we propose a multivariate copula-based approach for modeling and forecasting the realized volatility matrix, which can be considered as an extension of the bivariate copula approach. 

Our proposed approach can be generally described as follows. We first decompose the realized volatility matrices into Cholesky factors, and then utilize two methods to construct the multivariate copula: (1) we decompose the joint distribution of current Cholesky factors and their first lags into their marginal distributions and a multivariate copula function; (2) we decompose the joint distribution of the single element of current Cholesky factors and the first lags of the whole current Cholesky factors into their marginal distributions and a multivariate copula function. In addition, in order to compare the performance, we apply bivariate copulas to model each component of the Cholesky factors and also propose a copula-HAR combined model according to the dependency structure of empirical data. Same copula models are also applied to the log volatilities, another way to guarantee the positive-definiteness of the volatility matrix. To evaluate the performance of our model in practice, we compare the one-day ahead realized volatility matrix forecasts. We find that the multivariate copula-based approach can achieve statistical significance while bivariate copula-based approach can achieve economic significance.

The rest of the paper is structured as follows. In section \ref{sec:mod} we introduce the realized volatility matrix and describe the details of our modeling and forecasting procedures. Section \ref{sec:emp} presents the out-of-sample empirical results of our models both in terms of statistical precision and economical performance. Section \ref{sec:con} concludes the whole paper.

\section{Modeling and Forecasting Realized Volatility Matrix}
\label{sec:mod}
\subsection{Realized Volatility Matrix}
Suppose that $\mbS(t)=(S_{1}(t),\ldots,S_{p}(t))^{T}$ represents the prices of $p$ financial assets, which can be stated as follows:
\begin{equation*}
d\log\mbS(t)=\bolmu_{t}dt+\bolsigma_{t}^{T}d\mbB_{t},
\end{equation*}
where $\bolmu_{t}$ is a $p-$dimensional drift vector, $\mbB_t$ is a $p-$dimensional standard Brownian motion and $\bolsigma_t$ is a $p \times p$ matrix. Empirical results indicate that high-frequency data are suffered from microstructure noise \citep{zhang2005tale}. It is common to assume the observed log prices $\mbY(t)$ are contaminated by the microstructure noise, i.e.,
\[\mbY(t)=\textup{log}\mbS(t)+\bolepsilon(t),\]
where $\bolepsilon(t)$ is $i.i.d$ noise around the true prices and independent of $\log\mbS(t)$.

Let $\bolgamma(t)=\bolsigma_{t}^{T}\bolsigma_{t}$ be the spot volatility matrix of $\log\mbS(t)$. We are interested in the daily integrated volatility matrix, which is defined as follows: for day $t$,
\begin{equation*}
\bolGamma_{t} =\left ( \Gamma_{ij} \right )_{1\leq i,j\leq p}(t)=\int_{t-1}^{t}\bolgamma (s)ds=\int_{t-1}^{t}\bolsigma_{s}^{T}\bolsigma_{s}ds.
\end{equation*}
The integrated volatility matrix can be estimated with consistency using  realized volatility  based on high-frequency intra-day prices. Suppose we divide each trading day into $M$ intra-day periods, we define different time points as $0=\tau_0<\tau_1<\tau_2<\ldots<\tau_M=1$, then the $j$-th intra-day return for the $t$-th day can be calculated as
\begin{equation*}
\mbY_{j,t}=\mbY((t-1)+\tau_j)-\mbY((t-1)+\tau_{j-1}), j=1,\ldots,M.
\end{equation*}
When the sampling frequency is not super high,
\cite{barndorff2004econometric} defined the realized covariation matrix which is a consistent estimator of  $\bolGamma_{t}$ as 
\begin{equation}
\label{rv}
\hatbolGamma_{t}=\sum_{j=1}^{M}\mbY_{j,t}{\mbY'_{j,t}}.
\end{equation}
The realized covariance matrix are symmetric by construction and for $p<M$, positive definite almost surely. It can be further modified by reducing the microstructure noise \citep{zhang2005tale, zhang2006efficient, jacod2009microstructure}, taking nonsynchronicity  \citep{hayashi2005covariance, voev2007integrated, barndorff2011multivariate} and jumps \citep{christensen2010pre,  boudt2012jump} into account. In this paper, the realized volatility matrices are constructed by sampling from subgrids and taking the average, which is referred as the one-scale estimator in \cite{zhang2005tale}. This estimator is more robust than Eq. (\ref{rv}) to the market microstructure noise, and the non-synchronicity is mild under the chosen frequency  \citep{chiriac2011modelling}. One can also apply other integrated volatility estimator to our proposed methods stated below. 

\subsection{Modeling and Forecasting Realized Volatility Matrix Using Copula}
Appendix \ref{copula} includes an introduction about copula theory.
In this section, our primary goal is to build the dependence structure between consecutive observations of the integrated volatility matrix. The copula-based models have a conventional assumption that $\bolGamma_{t}$ is a Markov process  \citep{sokolinskiy2011forecasting, simard2015forecasting}. Thus we focus on the joint distribution of $\bolGamma_{t-1}$ and $\bolGamma_{t}$. 

To guarantee the positive definiteness of the volatility matrix forecasts, we
consider two methods. The first method is to apply Cholesky decomposition to $\hatbolGamma_{t}$, that is, there exists a matrix  $\mbP_t$ such that  $\mbP_t^{T}\mbP_t=\hatbolGamma_t$. Let $\mbX_{t}=\text{vech}(\mbP_t)$ which consists of the upper triangular components of $\mbP_t$, then $\mbX_{t}$ is a $m\times1$ vector where $m=\frac{n(n+1)}{2}$ and $n$ is the number of assets. Instead of applying copula models to $\hatbolGamma_t$, we will model $\mbX_{t}$. Another way to guarantee the positive definiteness is through matrix logarithm transformation. We employ this method and model the dynamics of the so-called log volatilities. Specifically, let $\mbA_{t}=\text{logm}(\widehat{\bolGamma}_{t})$, then $\mbA_{t}$ is a real, symmetric matrix, and the matrix exponential transformation performs a power series expansion, which will result in a real, semi-positive definite matrix $\widehat{\bolGamma}_{t}$, that is,
\begin{equation} \label{eq:log}
\widehat{\bolGamma}_{t}=\text{expm}(\mbA_{t})= \sum_{s=0}^{\infty }\left ( \frac{1}{s!} \right )\mbA_t^s.
\end{equation}
Denote $\mba_t=\text{vech}(\mbA_{t})$, we will also apply the copula models to the log-volatility series $\mba_t$. 

To construct copulas to model $\mbX_t$ process, we propose the following four methods\footnote{We only mention in details the methods for $\mbX_t$ to demonstrate the idea. One can apply the exact same procedure to the log-volatilities $\mba_t$.}.
\paragraph{Multivariate Copula Approach 1.}
Fit a $2m$-dimensional multivariate copula $C$ on $(\mbX_{t-1},\mbX_{t})$, by using which we can directly forecast $\mbX_{t+1}$. This approach allows us to model and forecast the entire matrix $\mbP_t$ by just using one multivariate copula.

\paragraph{Multivariate Copula Approach 2.}
Fit a $(m+1)$-dimensional multivariate copula $C$ on $(\mbX_{t-1}, X_{j,t})$ which allows us to get a directly forecast of $X_{j, t+1}$. By repeating this procedure for $j=1,\ldots,m$, we can get a forecast for $\mbX_{t+1}$. This approach can model and forecast each element of the matrix $\mbP_t$ individually.

\paragraph{Bivariate Copula Approach.} Fit a bivariate copula $C$ on $(X_{j, t-1},X_{j, t})$. By repeating this procedure for  $j=1,\ldots,m$, we can get a forecast for $\mbX_{t+1}$. This approach use less information from day $t-1$, and simply decompose $\mbX_t$ as $m$ univariate time series.

\paragraph{Copula-HAR Approach.}
Copula-based realized volatility model may outperform HAR \citep{sokolinskiy2011forecasting}. Empirical findings suggest correlations between the variances are higher than between the covariances in financial market. Therefore for the Cholesky decomposition matrix $\mbP_t$, we propose a combined approach: model the diagonal part with bivariate copulas, while model the covariances with HAR, and then obtain the forecast for $\mbX_{t+1}$.

The modeling and forecasting procedure of using bivariate copulas can be found in \cite{sokolinskiy2011forecasting}. For the multivariate case, without loss of generality, here we only present the procedure for the multivariate copula approach 1. This modeling process is developed from \cite{remillard2012copula} and \cite{simard2015forecasting}. Assume that $\mbX$ is Markovian, and ($\mbX_{t-1}$,$\mbX_{t}$) has continuous marginal distribution $\mbF$ and joint distribution $\mbH$. 
Let copula $Q(u)=C(\mbu,\mathbf{1})$, where $\mathbf{1}$ is a $m$-dimensional unit vector and $q$ represents its density, and $\mbU_t=\mbF(\mbX_{t})$, then by Eq. (\ref{eq: condition}) in Appendix \ref{copula}, the conditional copula of $\mbX_t$ given $\mbX_{t-1}$ is 
\begin{equation}
\label{eq:condition2}
C_{\mbU_{t}|\mbU_{t-1}}(\mbu_t|\mbu_{t-1})=\frac{\partial_{1}\ldots\partial_{m}C_{\mbU_{t-1},\mbU_{t}}(\mbu_{t-1},\mbu_{t})}{q_{\mbU_{t-1}}(\mbu_{t-1})}.
\end{equation}
This modeling process requires the estimation of the marginal distribution of $\mbX_{t}$. To avoid the misspecification, we estimate this marginal distribution $\mbF$ nonparametrically by using the empirical distribution function. Specifically, the estimate is defined as $\widehat{\mbF}=(\widehat{F}_1,\ldots,\widehat{F}_m)$, and 
\begin{equation}
\label{eq: empirical}
\widehat{F}_{j}(x)=\frac{1}{T+1}\sum_{t=1}^{T}I(X_{j,t}\leq x),\hspace{2mm} x \in\mathbb{R},j\in{1,\ldots,m},
\end{equation}
where $T$ denotes the sample size.

With the above modeling process, we can continue the forecasting procedure as follows. Suppose we have observations $\mbX_1,\ldots,\mbX_T$ and set $\mbX_T=\mby$, then the one-day ahead forecast of $\mbX_{T+1}$ can be obtained by
\begin{enumerate}
	\item Fit copula $C_{\mbU_t,\mbU_{t-1}}(\mbu_t,\mbu_{t-1})$ on the values of $(\mbX_{t-1},\mbX_{t})$, with $t=2,\ldots,T$. Calculate the conditional copula  $C_{\mbU_{t}|\mbU_{t-1}}(\mbu_t|\mbu_{t-1})$ of $\mbX_{t}$ given $\mbX_{t-1}$ by Eq. (\ref{eq:condition2}) and (\ref{eq: empirical}).
	\item Set $\mbu=\widehat{\mbF}(\mby)$, simulate $B$ realizations of the $\mbv^{(i)}=\widehat{\mbF}(\mbX_{T+1})$ from the fitted conditional copula  $\mbC_{\mbv|\mbu}(\mbv|\mbu)$, $i=1,2,..,B$.
	\item Use the inverse empirical distribution to transform each of the $B$ realizations into values of $\widehat{\mbX}_{T+1}$, i.e. $\widehat{\mbX}_{T+1}^{(i)}=\widehat{\mbF}^{-1}(\mbv^{(i)}), \vspace{2mm} i=1,\ldots,B$.
	\item $\widehat{\mbX}_{T+1}=\frac{1}{B}\sum_{i=1}^{B}{\widehat{\mbX}}_{T+1}^{(i)}$. Denote this forecast as $\widehat{\mbX}_{T+1|T}$.
\end{enumerate}
Then a positive-definite realized volatility matrix forecast $\widetilde{\bolGamma}_{T+1|T}$ can be obtained by 
$$\widetilde{\bolGamma}_{T+1|T} = \widehat{\mbP}_{T+1|T}^{T}\widehat{\mbP}_{T+1|T},\quad \vech\left(\widehat{\mbP}_{T+1|T}\right) = \widehat{\mbX}_{T+1|T}.$$
We can apply the same procedure to forecast $\mathbf{a}_{T+1}$, and obtain $\widetilde{\bolGamma}_{T+1|T}$ by matrix exponential transformation of $\widehat{\mba}_{T+1|T}$ using Eq. (\ref{eq:log}).

\subsection{Models for Comparison}
To evaluate the performance of our copula-based models, we employ the following three popular  models as benchmarks.
\paragraph{Heterogeneous Autoregressive (HAR) Model \citep{corsi2009simple}.}
The HAR model suggested an  AR-type model with the feature of considering volatilities averaged over different time horizons. Specifically, the series of each Cholesky element $X_{j,t}$ can be modeled as
\begin{equation*}
X_{j,t}=\beta_{0}+\beta_{d}X_{j,t-1}+\beta_{w}X_{j,t-1}^{(w)}+\beta_{m}X_{j,t-1}^{(m)}+\epsilon_{t,j},
\end{equation*}
where $X_{j,t-1}^{(w)}=\frac{1}{5}\sum_{l=0}^{4}X_{j,t-1-l}$, $X_{j,t-1}^{(m)}=\frac{1}{22}\sum_{l=0}^{21}X_{j,t-1-l}$, and $\epsilon_{t,j}$ is i.i.d Gaussian with mean 0 and variance $\sigma^2$. The coefficients can be easily estimated by OLS. 

\paragraph{Vector ARFIMA (VARFIMA) Model \citep{chiriac2011modelling}.}
A VARFIMA$(1,d,1)$ model will be used later in our empirical study as one of the benchmarks. It has the form
\begin{equation*}
(1-\phi L)\mbD(L)(\mbX_t-\mbc)=(1-\theta L)\bolepsilon_{t},\quad \bolepsilon_{t} \sim N(0,\bolSigma),
\end{equation*}
where $\mbc$ is an $m \times 1$ vector of constraints and $\mbD(L)=(1-L)^d\mbI_m$, $\phi$ and $\theta$ are scalars and $L$ is a lag operator with $L\cdot X_t = X_{t-1}$. This equation with restrictions on the AR, MA and fractionally integration operators can keep the estimation parsimonious. To fit this VARFIMA model, we use an extended version of approximate maximum likelihood approach proposed by \cite{beran1995maximum}. This estimation can effectively minimizes the residual sum of squares by avoiding the estimation of the $m\times m$ matrix $\bolSigma$. In practice the mean vector $\mbc$ is set to be the sample mean of $\mbX_t$. Forecasts can be obtained by the VMA($\infty$) and VAR($\infty$) representations \citep{chiriac2011modelling, lutkepohl2005new}.
\paragraph{Dynamic Conditional Correlation (DCC) Model \citep{engle2002dynamic}.}
 If we let $\mbr_t$ be a $n\times 1$ vector of log daily returns with $n$ being the number of assets, then process $\mbr_t$ can be written as $\mbr_t=E(\mbr_t| \mathcal{F}_{t-1})+\bolvarepsilon _t$, with $\bolvarepsilon_t=\mbH_t^{1/2}\mbz_{t}$, $E(\mbz_t)=0$, and $\textup{Cov}(\mbz_t)=\mbI_n$. $\mbH_t$ is the volatility matrix we are interested in.
If the conditional mean of daily return is assumed to be constant i.e. $E(\mbr_t| \mathcal{F}_{t-1})=\bolmu$, the DCC-GARCH can estimate the models on the demeaned series of daily returns. Specifically, the model is defined as
\begin{equation*}
\mbH_t=\mbD_t\mbR_t\mbD_t,\quad \mbD_t=\textup{diag}(d_{11,t}^{1/2},\ldots,d_{nn,t}^{1/2}),
\end{equation*}
where $d_{ii,t}=\omega_{i}+\alpha_{i}\varepsilon_{i,t-1}^2+\beta_id_{ii,t-1}$, $\omega_i,\alpha_i,\beta_i>0$, and  $\alpha_{i} + \beta_{i} < 1$. So $d_{ii,t}$ is a GARCH(1,1) process for $i=1,\ldots,n$. The dynamic correlation matrix is expressed as
\begin{equation*}
\mbR_t=(\textup{diag}(\mbQ_t))^{-1/2}\mbQ_t(\textup{diag}(\mbQ_t))^{-1/2},
\end{equation*}
where $\mbQ_t=(1-\theta_1-\theta_2)\overline{\mbQ}+\theta_1\mbu_{t-1}\mbu_{t-1}'+\theta_2\mbQ_{t-1}$,   $\mbu_t=(u_{t,1},\ldots,u_{t,n})$ with $u_{t,i}=\frac{\varepsilon_{i,t}}{\sqrt{d_{ii,t}}}$ and $\overline{\mbQ}$ is the unconditional covariance of $\mbu_t$. In terms of implementation of estimating and forecasting procedure of DCC-GARCH, we use the $\texttt{rmgarch}$ package in R.

\section{Empirical Study}
\label{sec:emp}
In this section, we will present an empirical application of our copula-based approaches for modeling and forecasting the realized volatility matrix. Models are fitted on the Cholesky factors $\mbX_t$ and log-volatilities $\mba_t$. We consider the following copula models:
\begin{itemize}
	\item Multivariate copula-based models
	\begin{itemize}
		\item Multivariate Student's $t$ copula (T-1) fit on $(\mbX_{t-1},\mbX_{t})$
		\item Multivariate Student's $t$ copula (T-2) fit on $(\mbX_{t-1},X_{j,t}),j=1,\ldots,m$
		\item Multivariate Clayton copula (CL-1) fit on $(\mbX_{t-1},\mbX_{t})$
		\item Multivariate Clayton copula (CL-2) fit on $(\mbX_{t-1},X_{j,t}),j=1,\ldots,m$
	\end{itemize}
	\item Bivariate copula-based models
	\begin{itemize}
		\item Bivariate Student's $t$ copula (Entry-T) fit on $(X_{j,t-1},X_{j,t}),j=1,\ldots,m$
		\item Bivariate Gumbel copula (Entry-GB) fit on $(X_{j,t-1},X_{j,t}),j=1,\ldots,m$
		\item Bivariate Clayton copula (Entry-CL) fit on $(X_{j,t-1},X_{j,t}),j=1,\ldots,m$
	\end{itemize}
	\item Bivariate copulas combined with HAR
	\begin{itemize}
		\item Student's $t$ copula with HAR (T-HAR)
		\item Gumbel copula with HAR (Gb-HAR)
		\item Clayton copula with HAR (Cl-HAR)
	\end{itemize}
\end{itemize}

\subsection{Data}
The data we use is obtained from Journal of Applied Econometrics Data Archive, and consists of tick-by-tick bid and ask quotes on stocks from NYSE. It contains the intra-day prices from 9:30 until 16:00 for the period January 1, 2000 to July 30, 2008 ($T=2156$ trading days) of six highly liquid stocks: American Express Inc. (AXP), Citigroup (C), General Electric (GE), Home Depot Inc. (HD), International Business Machines (IBM) and JPMorgan Chase \&Co. (JPM). 

For each day, 78 intraday returns can be obtained by sampling every 5 minutes. By Eq. (\ref{rv}), realized volatility matrix can be constructed by using these 5-minute returns. The estimator is further refined by a subsampling procedure by constructing 30 equal-spaced subgrids and computing the realized volatility matrix on each subgrid and taking the average. This estimator can help to reduce the effects from microstructure noise and non-synchronicity. We build models on both the Cholesky decomposition factors $\mathbf{X}_t$ and log-volatilities $\mathbf{a}_t$, whose summary statistics can be found in tables \ref{t3} and \ref{t4} in the Appendix \ref{tables}. The Cholesky factors exhibit the same characteristics as the realized volatility matrix elements, i.e., right skewed and leptokurtic. Also, the estimated Hurst Exponents in the tables indicate the long memory of the decomposed series.

With $T=2156$, we assess the performance of different models from an out-of-sample forecasting perspective. We choose a moving window of 1508 days (i.e., 6 years), with the first moving window  from January 1, 2000 to December 31, 2005. All models are re-estimated for each day in the moving windows and the corresponding one-day ahead forecast of realized volatility matrix are calculated.

\subsection{Performance Evaluation}
To evaluate the precision of statistical forecasting, we employ the root mean squared error (RMSE) criterion based on the Frobenius norm of matrix.  Denote $\widetilde{\bolGamma}_{t+1|t}$ is the realized volatility forecast, and $\widehat{\bolGamma}_{t+1|t}$ is the real realized volatility, then we have
\begin{eqnarray*}
\mbe_{t+1,t}&=&\widetilde{\bolGamma}_{t+1|t}-\widehat{\bolGamma}_{t+1|t}, \\
\textup{RMSE}&=&\frac{1}{T} \sum_{t=1}^{T}\sqrt{\sum_{i=1}^{n}\sum_{j\geq i} e_{t+1,t_{i,j}}^2}.
\end{eqnarray*}

To assess the economic value of volatility forecasts, we evaluate the portfolio optimization strategy proposed by  \cite{markowitz1952portfolio}. For a risk-averse investor with suitable utility function (for example second-degree polynomial or logarithmic), the portfolio optimization is equivalent to find the asset weights which minimizes the portfolio volatility $\widehat{\sigma}_t$ for a given expected return $\mu_p$. The optimal portfolio is given by solving the following quadratic problem
\begin{eqnarray*}
\textup{argmin}_{\bolomega_{t+1|t}}\sigma_{t+1|t}^p=\textup{argmin}_{\bolomega_{t+1|t}}\bolomega_{t+1|t}'{\widetilde{\bolGamma}}_{t+1|t}\bolomega_{t+1|t},\\
\textup{s.t.} \hspace{1mm} \bolomega_{t+1|t}'E_{t}(\mbr_{t:t+1})=\mu_{p} \hspace{1mm}\textup{and} \hspace{1mm} \bolomega_{t+1|t}'\mathbf{\ell}=1, \bolomega_{t+1|t}' \geq 0,
\end{eqnarray*}
where $\bolomega_{t+1|t}$ is the $n \times 1$ vector of portfolio weights chosen at $t$ for the period from $t$ to $t+1$, $\mathbf{\ell}$ is an $n\times 1$ vector of ones, and $\mbr_{t:t+1}$ is the ex-post portfolio return. By repeating this optimization for several levels of daily returns, we can compute different $r_{t+1|t}^p=\bolomega_{t+1|t}'\mbr_{t:t+1}$ and $\widehat{\sigma}_{t+1|t}^p=\sqrt{\bolomega_{t+1|t}'{\widehat{\bolGamma}}_{t+1|t}\bolomega_{t+1|t}}$. Then by averaging the minimal portfolio variance over all $t$, we obtain ex-post efficient frontiers for every forecasting model. 

More specifically, for a particular expected return $\mu_p$, we will find the point $(r_{t+1|t}^p, \widehat{\sigma}_{t+1|t}^p)$, which is corresponding to the  global minimum variance portfolio (GMVP). This optimization problem can be solved for different levels of $\mu_{p}$, thus given us an efficient frontier for each forecasting model. The efficient frontiers represent the best mean-variance trade-off portfolio that can be achieved by using the forecasts of different models. We also calculate the ideal  efficient frontier by using the ``oracle'' forecast, i.e., $\widetilde{\bolGamma}_{t+1|t}=\widehat{\bolGamma}_{t+1|t}$.

In addition, we apply the Model Confidence Set (MCS) methodology \citep{hansen2011model} to further evaluate the models. MCS is a set of models which contains the best model given a level of confidence. To obtain MCS, we start with the full set of candidate models $\mathcal{M}_0=\left \{1,..,m_0 \right \}$, where $m_0$ is the total number of models. For all models in the set, a loss differential between models will be computed based upon a loss function $L$, i.e., for $t=1,2,\ldots,T$ and model $i$ and $j$, $d_{ij,t}=L_{it}-L_{jt}$. We conduct the hypothesis $H_0: E[d_{ij}]=0$ for all $i,j \in \mathcal{M}$, and the test statistic is a range statistic which can be calculated as
\[T_{R,k}=\text{max}_{i,j \in \mathcal{M}}\left | t_{ij} \right |=\text{max}_{i,j \in \mathcal{M}}\frac{\overline{d}_{ij}}{\sqrt{\widehat{\text{var}}(\overline{d}_{ij})}}, \quad \overline{d}_{ij}=\frac{1}{T}\sum_{t=1}^{T}d_{ij}, \]
where $\widehat{\text{var}}(\overline{d}_{ij})$ is obtained from a block-bootstrap procedure. If $H_{0}$ is rejected at a given significance level $\alpha$, the worst model is then removed from the set. This procedure will be repeated until no model to be removed from the set. This MCS method allows us to compare models without benchmarks. 

In our evaluation of statistical precision, we will use the Stein Loss function \citep{james1961estimation} defined as $L(\mathbf{Y}_t,\widehat{\mathbf{Y}}_t)=\tr(\widehat{\mathbf{Y}}_t\mathbf{Y}_t^{-1})-\textup{ln}|\widehat{\mathbf{Y}}_t\mathbf{Y}_t^{-1}|-N$, which is also called as multivariate quasi likelihood (MVQLIKE). According to \cite{laurent2013loss}, Stein Loss is consistent in the sense that (1) it can preserve the true ranking of the covariance models and (2) it punishes more heavily on underpredictions. For economic comparison, we will use MCS to select the set of models which contains the one with the smallest standard deviation at 5\% confidence level. 

\subsection{Out-of-Sample Forecasting Results -- Statistical Evaluation}
We first report the out-of-sample forecasting RMSEs for 648 days in table \ref{t1}, from which we can find that HAR has the smallest RMSE among all models then followed by VARMIFA , T-HAR, Gb-HAR and two multivariate-$t$ copulas models. Under 5\% level of confidence, HAR, VARFIMA and T-1, T-2 all belong to the MCS by using Stein loss function. This indicates that the multivariate copula-based models can obtain statistical significance. This conclusion holds for both Cholesky factors and log-volatilities. In addition we notice that the matrix logarithm transformation in general will give a higher RMSE than the Cholesky decomposition method. This is potentially caused by the procedure of retransformation forecasts of $\widehat{\Gamma}_{t}$, which will be naturally biased by Jensen's inequality. 
\begin{table}[!htbp]

	\centering
	\caption{RMSEs of out-of-sample forecast for 648 days (window size = 1508 days). The bold RMSEs represent models in the 5\% MCS.}
		\label{t1}
	\begin{tabular}{cc|cc}
		\hline\hline
		\multicolumn{2}{c|}{Model}  &       Cholesky Factor&     Log Volatility\\\hline
		& DCC-GARCH&    5.0918  &                          5.0918\\
		Benchmark & HAR&            $\mathbf{3.9263}$&                             $\mathbf{4.0251}$\\
		& VARFIMA&    $\mathbf{3.9799}$&                             $\mathbf{4.0611}$\\
		\hline
		& T-1&             $\mathbf{4.1959}$&                                 $\mathbf{4.3062}$\\
		Multivariate & T-2&            $\mathbf{4.1871}$&                                  $\mathbf{4.3047}$\\     
		Models & CL-1&         6.2199&                                 5.9004\\
		& CL-2&          6.3734&                                5.8714\\
		\hline
		Bivariate & Entry-T&       4.4006&                                4.8001\\
		Models & Entry-GB&    4.3316&                                4.5117\\
		& Entry-CL&     5.5547&                                5.7369\\
		\hline
		Bivariate & T-HAR&       4.0812&                                4.3943\\
		+ HAR & Gb-HAR&    4.1044&                                4.3318\\
		 & Cl-HAR&      4.3716&                                5.2523\\\hline
		\hline
	\end{tabular}
\end{table}

We then focus on the bivariate copula combined with HAR modeling approach.  The Cholesky matrix is labeled as below (the log-volatility matrix is labeled in the same way):
\begin{equation*}
\mathbf{P}_{t}=\begin{pmatrix}
1&\quad  2&\quad 4 &\quad 7 &\quad 11 &\quad 16&\\ 
&\quad  3&\quad  5&\quad  8&\quad  12&\quad 17&\\ 
&\quad  &\quad  6&\quad  9&\quad  13&\quad 18&\\ 
&\quad  &\quad  &\quad  10&\quad  14&\quad 19&\\ 
&\quad  &\quad  &\quad  &\quad  15&\quad 20&\\ 
&\quad  &\quad  &\quad  &\quad  &\quad 21& 
\end{pmatrix}
\end{equation*}
To check the correlation patterns, we compute the rank correlation coefficient for the $(\mathbf{X}_{t-1},\mathbf{X}_{t})$ and $(\mathbf{a}_{t-1},\mathbf{a}_{t})$. The correlation matrices are shown in the figure \ref{heatmap} below.
\begin{figure}[!htb] 
	\minipage{0.5\textwidth}
	\includegraphics[width=\linewidth,height=7cm]{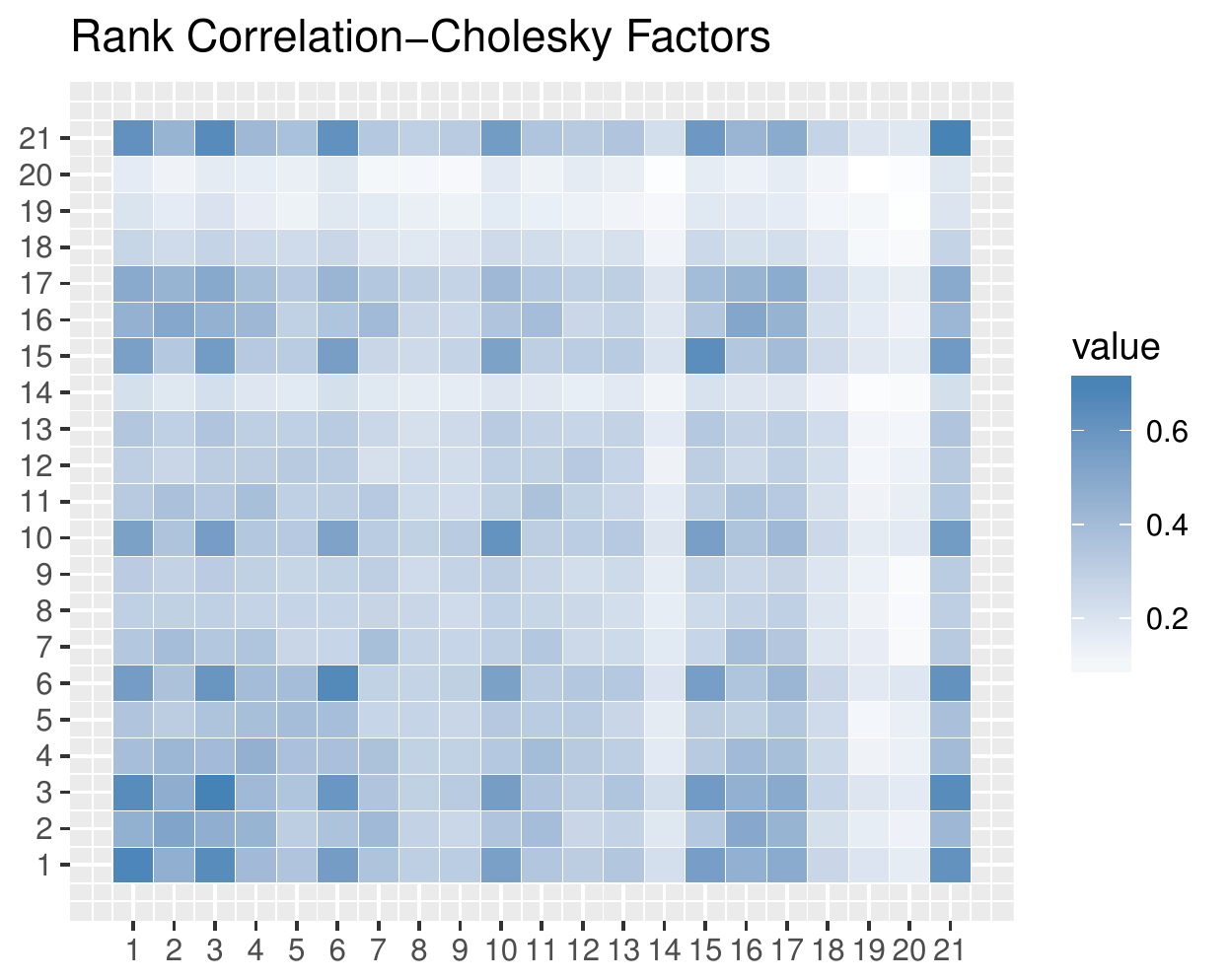}
	\raggedright\centering{\footnotesize{Cholesky Factors}}
	\endminipage\hfill
	\minipage{0.5\textwidth}
	\includegraphics[width=\linewidth,height=7cm]{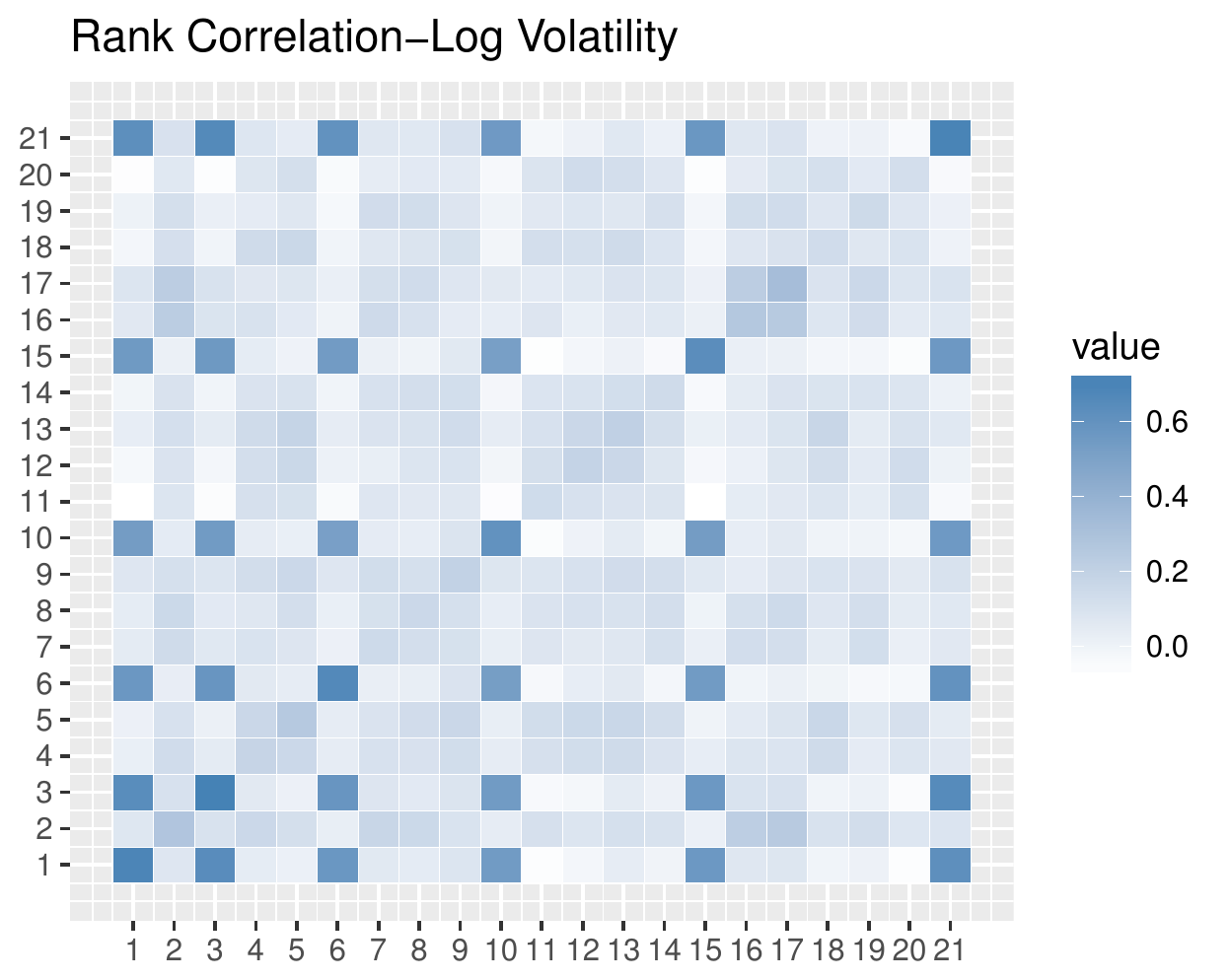}
	\raggedright\centering{\footnotesize{Log Volatilities}}
	\endminipage
	\centering
	\caption{Heatmaps of the rank correlations.} \label{heatmap}
\end{figure}

There exists a clear higher dependence between the diagonal entries of Cholesky factors and log-volatility matrices (i.e., nodes 1, 3, 6, 10, 15, and 21), which is visualized by the corresponding cells being darker. This pattern supports the phenomenon that correlations between the variances are larger than those between the covariances. Compared with the original bivariate copula models, the RMSEs of the HAR combined models are reduced due to the HAR component. Among these three combined models, the T-HAR model outperforms the other two for Cholesky factors, which suggests that an equal positive tail dependence structure for the diagonal part of Cholesky factors may help to obtain accurate forecasts. For the log-volatility, the Gb-HAR has the smallest RMSE, which suggests that a positive upper tail dependence structure for the diagonal part of log-volatility matrix may be helpful for getting more precise forecasts.

\subsection{Out-of-Sample Forecasting Results -- Economic Evaluation}
Table \ref{t2} contains the average of the realized conditional standard deviation of the global minimum variance portfolio (GMVP). We use the MCS methodology to select the set of models which contains the model with the smallest standard deviation at the 5\% confidence level.  

\begin{table}[!htp]
	
	\centering
	\caption{Annualized conditional standard deviation of the GMVP (window size = 1508 days). The bold SDs represent models in the 5\% MCS.}
	\label{t2}
	\begin{tabular}{cc|cc}
		\hline\hline
		\multicolumn{2}{c|}{Model} &      Cholesky Factor&    Log Volatility\\\hline
		& DCC-GARCH&      13.0286&   13.0286   \\
		Benchmark & HAR&                     $\mathbf{12.5694}$&          $\mathbf{12.5715}$                  \\
		& VARFIMA&    12.6646&                           $\mathbf{12.6679}$\\
		\hline
		& T-1&             12.7400&                      12.7387          \\
		Multivariate & T-2&           12.7430&                        12.7409        \\    
		Models & CL-1&         13.3386&                       13.2607       \\
		& CL-2&          13.3153&                       13.2156         \\
		\hline
		Bivariate & Entry-T&     12.8568&                         12.8721     \\
		Models & Entry-GB&   12.7939&                         12.8142     \\
		& Entry-CL&     13.1412&                          13.0997    \\
		\hline
		Bivariate & T-HAR&         $\mathbf{12.6816}$ &      12.7359                        \\
		+ HAR & Gb-HAR&       $\mathbf{12.6610}$&       12.7120                      \\
		& Cl-HAR&         12.7535&                      12.8408        \\\hline
		\hline
	\end{tabular}
\end{table}

For the 13 models we discussed in the paper, the economic evaluation results are very similar to the model forecast RMSEs, that is, HAR, VARFIMA, T-HAR, Gb-HAR, T-1 and T-2 models are the ones with the smallest standard deviations for both Cholesky factors and log-volatilities. Under the Cholesky decomposition method, HAR, T-HAR and Gb-HAR models belong to the 5\% MCS, while only HAR and VARFIMA are selected to the 5\% MCS by using the matrix logarithm transformation.

We further look at the efficient frontiers of the models, with the results of the Cholesky factors and log-volatilities given in figures \ref{fig2} and \ref{fig3} (in Appendix \ref{tables}), respectively. Obviously, the oracle forecast leads to the best mean-variance trade-off. Among the 13 models we discussed, the benchmarks HAR and VARFIMA have the best efficient frontiers, T-1, T-2, T-HAR, Gb-HAR are the second best models, Cl-HAR, Entry-GB and Entry-T are the third best, while DCC, Entry-Cl, and Cl-1, Cl-2 have the worst performance. 

\begin{figure}[!htb]
	\minipage{0.5\textwidth}
	\includegraphics[width=\linewidth,height=6cm]{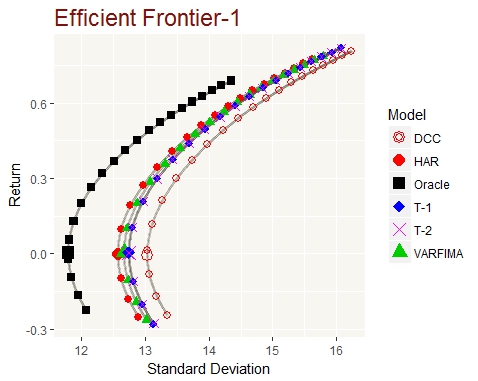}
	\endminipage\hfill
	\minipage{0.5\textwidth}
	\includegraphics[width=\linewidth,height=6cm]{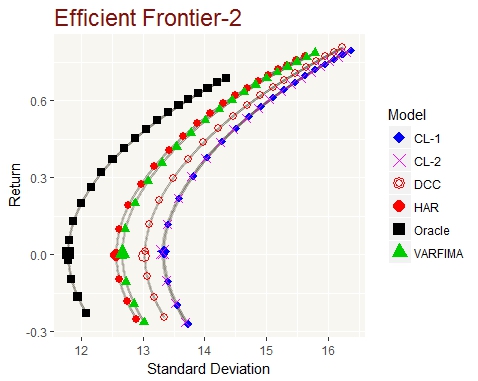}
	\endminipage \\
		\minipage{0.5\textwidth} 
	\includegraphics[width=\linewidth,height=6cm]{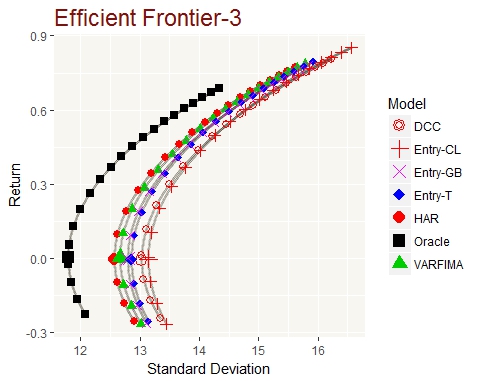}
	\endminipage\hfill
	\minipage{0.5\textwidth}
	\includegraphics[width=\linewidth,height=6cm]{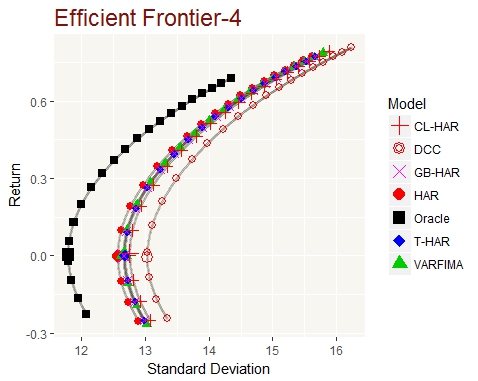}
	\endminipage
	\caption{Portfolio efficient frontiers using Cholesky factors. The results of oracle and three benchmarks are included in all four graphs for a better comparison.}
	\label{fig2}
\end{figure}

Economic evaluation is indeed a different criteria than RMSE that can help us to evaluate the copula models. Under different evaluation criterion, the copula-based forecasting models can always achieve certain significance results, such as the T-HAR and Gb-HAR models for the Cholesky factors. 

\subsection{Discussion: about Moving Window Size} 
In order to test the sensitivity of the moving window size, we repeat our analysis by using a different moving window of 1000 days, with the results summarized in table \ref{t6}. In this case, the data are split into an in-sample with 1000 days and out-of-sample with 1156 days.

\begin{table}[!htp]
	\centering
	\caption{Evaluation results for moving window = 1000 days. The bold numbers represent models in the 5\% MCS.}
	\label{t6}
	\begin{tabular}{cc|cc|cc}
		\hline\hline
		 & &      \multicolumn{2}{c|}{Cholesky Factor} &  \multicolumn{2}{c}{Log Volatility} \\		\multicolumn{2}{c|}{Model} & RMSE & SD & RMSE & SD \\
		\hline
		& DCC-GARCH&      3.5230& 11.7414& 3.5230&   11.7414           \\
		Benchmark & HAR&                  $ \mathbf{2.7610}$& $\mathbf{11.2001}$&  $\mathbf{2.7944}$&   $\mathbf{11.1958}$                          \\
		& VARFIMA&   $\mathbf{2.8323}$&  $\mathbf{11.2640}$&  $\mathbf{2.8656}$& $\mathbf{11.2818}$                           \\
		\hline
		& T-1&             $\mathbf{3.0498}$& 11.3232& $\mathbf{3.1317}$&  11.3165                              \\
		Multivariate & T-2&          $\mathbf{3.0445}$& 11.3212&  $\mathbf{3.1276}$&   11.3162                              \\  
		Models & CL-1&         3.9394& 11.7775&  4.3776& 11.7746                             \\
		& CL-2&          3.8782& 11.7505&  4.3138& 11.7359                           \\
		\hline
		Bivariate & Entry-T&     3.4691& 11.4196&3.6533 &  11.4369                    \\
		Models & Entry-GB&   3.1938& 11.3623& 3.3014& 11.3722                         \\
		& Entry-CL&     4.3699& 11.7238& 4.3162& 11.7475                       \\
		\hline
		Bivariate & T-HAR&        3.1273& 11.4003&  3.3756& 11.3418                            \\
		+ HAR & Gb-HAR&     2.9641& 11.3455&   3.1334& 11.3069                            \\
		& Cl-HAR&       3.6025& 11.6182&  4.0880&  11.5243                           \\\hline
		\hline
	\end{tabular}
\end{table}

By reducing the moving window size to 1000 days, there is  an overall decrease of the RMSEs and standard deviations of GMVP. This is potentially because that we now have a larger portion of good forecasting results in the out-of-sample forecasting period. When the dates are away from year 2008, the market has a good economic condition, and thus the volatilities are stable and have relative small values. As a result, the corresponding forecasting errors will be smaller. Other than that, table \ref{t6} shows a similar pattern as the results with moving window size = 1508 days. 

\section{Conclusion}
\label{sec:con}
In this paper, we proposed a copula-based approach to model the dynamics of realized volatility matrices, and forecast their future values. The models can explicitly capture the hidden dependence structure of the realized volatility matrix. To guarantee the positive definiteness of the volatility matrix forecasts, the volatility matrices are decomposed into Cholesky factors or transformed through matrix logarithm. After the decomposition/transformation, they are further modeled by different multivariate and bivariate copulas. 

In an empirical application, we evaluated the forecasting results not only in terms of statistical comparison, but also in terms of improving the performance of mean-variance efficient portfolios. Our copula-based models can be selected in the model confidence sets and thus achieve the significance. In addition, we compared the results under different window sizes. Although the exact numbers may differ, the patterns among models keep the same for different window sizes.

\bibliographystyle{ims}
\bibliography{bi}

\appendix
\section{Appendix}
\subsection{The Copula Theory} \label{copula}
In this appendix we introduce some basic knowledge about copulas, where more details can be found in \cite{nelsen2007introduction}.
\begin{definition}
	$C$:$[0,1]^{d}\rightarrow [0,1]$ is a d-dimensional copula if $C$ is a joint distribution function of a d-dimensional random vector on the domain $[0,1]^{d}$ with uniform margins.
\end{definition}
More specifically, consider a random vector $(X_{1},X_{2},\ldots,X_{d})$, if the marginal distribution $F_{i}(x)=P(X_{i}\leq x)$ is continuous, then the random vector $(U_{1},\ldots, U_{d})=(F_{1}(X_{1}),\ldots,F_{d}(X_{d}))$ is uniformly distributed. The copula of $(X_{1},X_{2},\ldots,X_{d})$ is then defined as the joint cumulative distribution function of $(U_{1},U_{2},\ldots,U_{d})$. For any $\textbf{u}=(u_1,\ldots,u_d) \in [0,1]^d$,
\begin{equation*}
C(u_{1},u_{2},\ldots,u_{d})=P(U_{1}\leq u_{1}, U_{2} \leq u_{2},\ldots,U_{d} \leq u_{d}).
\end{equation*}
Sklar's theorem provides the theoretical foundation for most statistics applications of copulas.
\begin{theorem}
	\label{sklar}
	\emph{(Sklar's Theorem, 1959)}
	Let $\bmH$ be a $d$-dimensional distribution function with margins $F_{1}$,\ldots,$F_{d}$. Then there exists a $d$-dimensional copula $C$ such that for all $(x_{1},\ldots,x_{d}) \in \overline{R}^d$, \[H(x_1,\ldots,x_d)=C(F_{1}(x_1),\ldots,F_{d}(x_{d})).\]
\end{theorem}
Followed by the Sklar's Theorem, assuming the marginal density $f_j$ of $F_j$ exists for each $j=1,\ldots,d$, then the joint density of $(X_1,\ldots,X_d)$ is given by
\[h(x_1,\ldots,x_d)=c((F_1(x_1),\ldots,F_d(x_d))\prod_{j=1}^{d}f_j(x_j),\]
where $c(\cdot)=\frac{\partial^dC\left ( \cdot \right )}{\partial u_1\ldots\partial u_d}$ is the density of copula function $\rm{\textit{C}}$. Let $d=d_1+d_2$, then the conditional density of $(X_{d_1+d_2}, X_{d_1+d_2-1},\ldots,X_{d_1+1})$ given $(X_1,\ldots,X_{d_1})$ is given by
\begin{equation*}
h(x_{d_1+d_2}, x_{d_1+d_2-1},\ldots,x_{d_1+1}|x_1,\ldots,x_{d_1})=\prod_{j=d_1+1}^{d}f_j(x_j)\frac{c(F_1(x_1),\ldots,F_d(x_{d}))}{c(F_1(x_1),\ldots,F_{d_1}(x_{d_1}))}.
\end{equation*}
This representation shows that we can separate the dependence structure from the marginal distribution. By choosing different copulas, we imply different dependence structure for the volatility.

While there exists a wide range of bivariate parametric copulas available to choose \citep{nelsen2007introduction}, the types of multivariate copulas are limited. For bivariate case, we have elliptical copulas including Gaussian and Student's $t$ copulas, and Archimedean copulas  such as Clayton, Gumbel copulas. The multivariate extensions commonly used in practice are Student's $t$ and Clayton copula. Different bivariate copulas can be compared by different tail dependence structures. Gaussian copula is tail independent while Student's $t$ has symmetric lower and upper tail dependence. Clayton copula has zero upper tail dependence and positive lower tail dependence while Gumbel copula has zero lower tail dependence and positive upper tail dependence. 

We now introduce conditional copula, which is crucial for our modeling and forecasting procedure \citep{patton2006modelling}. Suppose $\mbX$ is a $d_1$-dimensional random vector with marginal distribution $\mbH_1=(F_1,\ldots,F_{d_1})$ and set $\mbU=\mbH_1(\mbX)=(F_1(X_1),\ldots,F_{d_1}(X_{d_1}))$; $\mbY$ is a $d_2$-dimensional random vector with marginal distribution $\mbH_2=(F_{d_1+1},\ldots,F_{d_1+d_2})$; denote $\mathbf{V}=\mathbf{H}_2(\mathbf{Y})=\left (F_{d_1+1}(Y_{1}),\ldots,F_{d_1+d_2}
(Y_{d_2})\right)$, and the joint distribution of $(\mathbf{X},\mathbf{Y})$ as $\mathbf{H}$ with density $\mathbf{h}$. The conditional distribution of $\mbY|\mbX$ can be obtained as follows:
\begin{equation}
\label{eq: condition}
F_{\mbY|\mbX}(\mby|\mbx)=C_{\mbV|\mbU}(\mbv|\mbu):=\frac{\partial^{d_1} C_{\mbU,\mbV}(\mbu,\mbv)}{\partial U_1 \cdots\partial U_{d_1}}=\frac{\partial_{u_1}\cdots\partial_{u_{d_1}}C_{\mbU,\mbV}(\mbu,\mbv)}{c_{\mbU}(\mbu)},
\end{equation}
with density $c_{\mbV|\mbU}(\mbv|\mbu)=\frac{c_{\mbU,\mbV}(\mbu,\mbv)}{c_{\mbU}(\mbu)}$ and $c_{\mbU}$ is the density of the copula $C_{\mbU}(\mbu)=C_{\mbU, \mbV}(\mbu,1,\ldots,1)$.
\newpage

\subsection{Supplementary Tables and Figures} \label{tables}
	\begin{table}[!htp]
		\centering
		\caption{Descriptive statistics of Cholesky elements.}
		\label{t3}
		\begin{tabular}{p{2.5cm} p{1.5cm} p{1.5cm} p{1.5cm} p{1.5cm} p{1.5cm} p{1.5cm} p{1.5cm}}
			\hline\hline
			Cholesky  Factors & Mean & Max & Min & Std & Skewness & Kurtosis & Hurst-Exponent\\\hline
			AXP&	1.5894&	7.5884	&0.2708	&0.9576	&1.4610&	6.3760&	0.9196\\\hline
			C&	1.4091&	8.6886	&0.3222	&0.8231	&1.8194	&9.7476&	0.9171\\\hline
			GE&	1.1729&	5.6248&	0.3192&	0.6114&	1.4868&	6.3342&	0.9510\\\hline
			HD&	1.4582&	5.7538	&0.3795&	0.6721&	1.5699&	6.6878&	0.9117\\\hline
			IBM&	1.1227	&4.7383	&0.2382&	0.5991&	1.7808&	6.9016&	0.9356\\\hline
			JPM&	1.3213	&9.0164	&0.2990&	0.7740&	2.1136&	13.5257&	0.9268\\\hline
			AXP-C&	0.7121&	7.6275&	-0.2129&	0.6640&	2.8298&	16.7133&	0.8470\\\hline
			AXP-GE&	0.5220&	3.9912&	-0.5047&	0.4307&	2.1730&	11.4125&	0.8754\\\hline
			AXP-HD&	0.5468	&3.7090&	-0.7045&	0.4632&	2.0120&	9.4415&	0.8264\\\hline
			AXP-IBM& 	0.4480	&3.0881&	-0.4310&	0.3454&	1.9625&	9.9524&	0.8410\\\hline
			AXP-JPM&	0.7151	&8.1623&	-0.2910&	0.6558&	2.8841&	18.5482&	0.8472\\\hline
			C-GE& 	0.3664	&2.6837 &-0.2848	&0.2956	&1.9918	&9.6918	&0.8654\\\hline
			C-HD& 	0.3569	&2.3988&	-0.3769&	0.2986&	1.7698&	8.7879&	0.8175\\\hline
			C-IBM	& 0.3129	&4.6760	&-1.2171	&0.2713&	3.4193&	39.3578&	0.8545\\\hline
			C-JPM&	0.5823&	5.7675&	-0.1875&	0.4329	&2.2633	&15.5025&	0.8771\\\hline
			GE-HD&	0.2846	&2.7585&	-0.5343&	0.2603&	1.6882&	10.2956&	0.8057\\\hline
			GE-IBM&	0.2575	&2.1344&	-0.4326&	0.2191&	1.8205&	10.2040&	0.8417\\\hline
			GE-JPM&	0.2184&	2.2185&	-0.3567&	0.2245&	1.8627&	10.2029&	0.8378\\\hline
			HD-IBM&	0.1523	&1.7051&	-0.4824&	0.1657&	1.3610&	10.5253&	0.8273\\\hline
			HD-JPM&	0.1364&	1.1899&	-0.9082&	0.1846&	0.6269&	6.6289&	0.8181\\\hline
			IBM-JPM	& 0.1352	&1.3738&	-0.8742&	0.1887&	1.2512&	8.4828&	0.8275\\\hline
			\hline
		\end{tabular}
	\end{table}

	\begin{table}[!htp]
	
		\centering
		\caption{Descriptive statistics of log-volatility elements.}
			\label{t4}
		\begin{tabular}{p{2.5cm} p{1.5cm} p{1.5cm} p{1.5cm} p{1.5cm} p{1.5cm} p{1.5cm} p{1.5cm}}
			\hline\hline
			Log-volatilites & Mean & Max & Min & Std & Skewness & Kurtosis & Hurst-Exponent\\\hline
			AXP	&0.3215	&3.6330	&-2.7222	&1.1299	&0.1337	&2.1925	&0.9571\\\hline
			C	&0.2953	&4.3465	&-2.2802	&1.0702	&0.3306	&2.3614	&0.9489\\\hline
			GE	&0.0925	&3.5218	&-2.3590	&0.9597	&0.3032	&2.4299	&0.9633\\\hline
			HD	&0.6486	&3.6313	&-2.0131	&0.8420	&0.3451	&2.7710	&0.9260\\\hline
			IBM	&0.0997	&3.5094	&-2.2060	&0.9130	&0.5709	&2.9327	&0.9426\\\hline
			JPM	&0.4682	&4.8485	&-2.3544	&1.0694	&0.2532	&2.4217	&0.9494\\\hline
			AXP-C	&0.3021	&0.9408	&-0.3468	&0.1582	&0.2619	&3.3190	&0.8446\\\hline
			AXP-GE	&0.2407	&0.6685	&-0.2366	&0.1361	&-0.0770	&3.0622	&0.8529\\\hline
			AXP-HD	&0.2048	&0.7330	&-0.3392	&0.1349	&0.0387	&3.0808	&0.8358\\\hline
			AXP-IBM	&0.2094	&0.6211	&-0.3406	&0.1297	&-0.1845	&3.2712	&0.8596\\\hline
			AXP-JPM	&0.2894	&0.9937	&-0.1551	&0.1625	&0.5539	&3.6991	&0.8909\\\hline
			C-GE	&0.2744	&0.7124	&-0.2412	&0.1403	&-0.0165	&2.8976	&0.8506\\\hline
			C-HD	&0.2161	&0.6557	&-0.2541	&0.1317	&0.0665 &3.0002	&0.8517\\\hline
			C-IBM	 &0.2295	&0.7953	&-0.2831	&0.1286	&-0.0383	&3.1668	&0.8652\\\hline
			C-JPM	&0.4195	&1.1256	&-0.1061	&0.1804	&0.5048	&3.4976	&0.8794\\\hline
			GE-HD	&0.2247	&0.6646	&-0.2719	&0.1340	&-0.0911	&3.1321	&0.8264\\\hline
			GE-IBM	&0.2582	&0.6679	&-0.1928	&0.1377	&-0.0557	&2.7872	&0.8412\\\hline
			GE-JPM	&0.2333	&0.6316	&-0.3160	&0.1302	&0.0182	&2.9692	&0.8063\\\hline
			HD-IBM	&0.2074	&0.6481	&-0.3070	&0.1298	&-0.0838	&3.1797	&0.8543\\\hline
			HD-JPM	&0.2024	&0.7488	&-0.3180	&0.1336	&0.0473	&3.2501	&0.8601\\\hline
			IBM-JPM	&0.2094	&0.6172	&-0.3279	&0.1281	&0.0070	&3.0698	&0.8051\\\hline
			\hline
		\end{tabular}
	\end{table}
	
	\begin{figure}[!htb]
		\minipage{0.5\textwidth}
		\includegraphics[width=\linewidth,height=6.5cm]{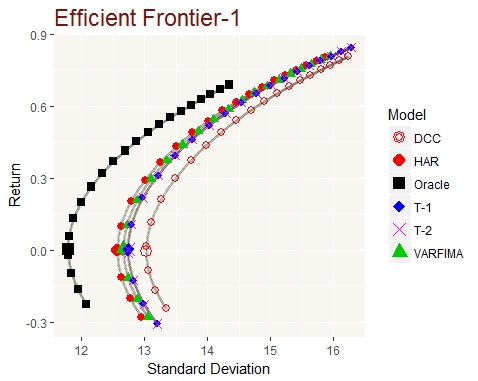}
		\endminipage\hfill
		\minipage{0.5\textwidth}
		\includegraphics[width=\linewidth,height=6.5cm]{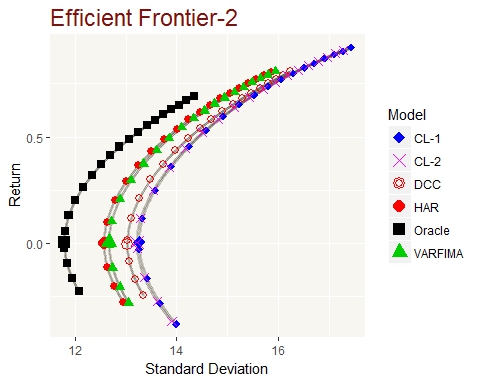}
		\endminipage \\

		\minipage{0.5\textwidth}
		\includegraphics[width=\linewidth,height=6.5cm]{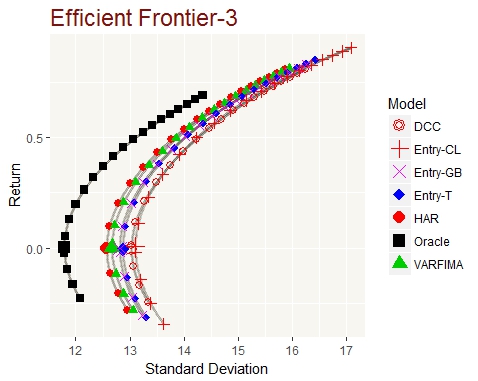}
		\endminipage\hfill
		\minipage{0.5\textwidth}
		\includegraphics[width=\linewidth,height=6.5cm]{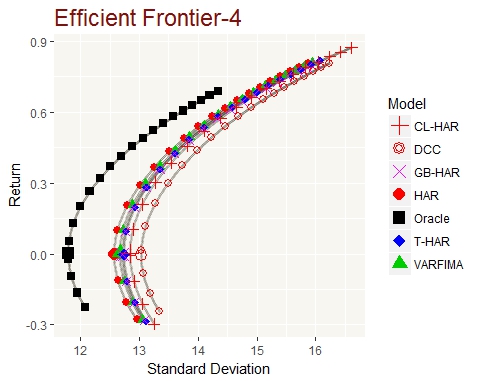}
		\endminipage
		\caption{Portfolio efficient frontiers using logarithm transformation. The results of oracle and three benchmarks are included in all four graphs for a better comparison.} \label{fig3}
	\end{figure}


\end{document}

%% file: setting.tex
	
	{
		\setlength{\baselineskip}{1.75\baselineskip} 
		
		\global\long\def\mba{\mathbf{a}}
		\global\long\def\mbA{\mathbf{A}}
		\global\long\def\mbb{\mathbf{b}}
		\global\long\def\mbB{\mathbf{B}}
		\global\long\def\mbc{\mathbf{c}}
		\global\long\def\mbC{\mathbf{C}}
		\global\long\def\mbd{\mathbf{d}}
		\global\long\def\mbD{\mathbf{D}}
		\global\long\def\mbe{\mathbf{e}}
		\global\long\def\mbE{\mathbf{E}}
		\global\long\def\mbf{\mathbf{f}}
		\global\long\def\mbF{\mathbf{F}}
		\global\long\def\mbg{\mathbf{g}}
		\global\long\def\mbG{\mathbf{G}}
		\global\long\def\mbh{\mathbf{h}}
		\global\long\def\mbH{\mathbf{H}}
		\global\long\def\mbi{\mathbf{i}}
		\global\long\def\mbI{\mathbf{I}}
		\global\long\def\mbj{\mathbf{j}}
		\global\long\def\mbJ{\mathbf{J}}
		\global\long\def\mbk{\mathbf{k}}
		\global\long\def\mbK{\mathbf{K}}
		\global\long\def\mbl{\mathbf{l}}
		\global\long\def\mbL{\mathbf{L}}
		\global\long\def\mbm{\mathbf{m}}
		\global\long\def\mbM{\mathbf{M}}
		\global\long\def\mbn{\mathbf{n}}
		\global\long\def\mbN{\mathbf{N}}
		\global\long\def\mbo{\mathbf{o}}
		\global\long\def\mbO{\mathbf{O}}
		\global\long\def\mbp{\mathbf{p}}
		\global\long\def\mbP{\mathbf{P}}
		\global\long\def\mbq{\mathbf{q}}
		\global\long\def\mbQ{\mathbf{Q}}
		\global\long\def\mbr{\mathbf{r}}
		\global\long\def\mbR{\mathbf{R}}
		\global\long\def\mbs{\mathbf{s}}
		\global\long\def\mbS{\mathbf{S}}
		\global\long\def\mbt{\mathbf{t}}
		\global\long\def\mbT{\mathbf{T}}
		\global\long\def\mbu{\mathbf{u}}
		\global\long\def\mbU{\mathbf{U}}
		\global\long\def\mbv{\mathbf{v}}
		\global\long\def\mbV{\mathbf{V}}
		\global\long\def\mbw{\mathbf{w}}
		\global\long\def\mbW{\mathbf{W}}
		\global\long\def\mbx{\mathbf{x}}
		\global\long\def\mbX{\mathbf{X}}
		\global\long\def\mby{\mathbf{y}}
		\global\long\def\mbY{\mathbf{Y}}
		\global\long\def\mbz{\mathbf{z}}
		\global\long\def\mbZ{\mathbf{Z}}

		\global\long\def\hatmba{\widehat{\mathbf{a}}}
		\global\long\def\hatmbA{\widehat{\mathbf{A}}}
		\global\long\def\hatmbb{\widehat{\mathbf{b}}}
		\global\long\def\hatmbB{\widehat{\mathbf{B}}}
		\global\long\def\hatmbc{\widehat{\mathbf{c}}}
		\global\long\def\hatmbC{\widehat{\mathbf{C}}}
		\global\long\def\hatmbd{\widehat{\mathbf{d}}}
		\global\long\def\hatmbD{\widehat{\mathbf{D}}}
		\global\long\def\hatmbe{\widehat{\mathbf{e}}}
		\global\long\def\hatmbE{\widehat{\mathbf{E}}}
		\global\long\def\hatmbf{\widehat{\mathbf{f}}}
		\global\long\def\hatmbF{\widehat{\mathbf{F}}}
		\global\long\def\hatmbg{\widehat{\mathbf{g}}}
		\global\long\def\hatmbG{\widehat{\mathbf{G}}}
		\global\long\def\hatmbh{\widehat{\mathbf{h}}}
		\global\long\def\hatmbH{\widehat{\mathbf{H}}}
		\global\long\def\hatmbi{\widehat{\mathbf{i}}}
		\global\long\def\hatmbI{\widehat{\mathbf{I}}}
		\global\long\def\hatmbj{\widehat{\mathbf{j}}}
		\global\long\def\hatmbJ{\widehat{\mathbf{J}}}
		\global\long\def\hatmbk{\widehat{\mathbf{k}}}
		\global\long\def\hatmbK{\widehat{\mathbf{K}}}
		\global\long\def\hatmbl{\widehat{\mathbf{l}}}
		\global\long\def\hatmbL{\widehat{\mathbf{L}}}
		\global\long\def\hatmbm{\widehat{\mathbf{m}}}
		\global\long\def\hatmbM{\widehat{\mathbf{M}}}
		\global\long\def\hatmbn{\widehat{\mathbf{n}}}
		\global\long\def\hatmbN{\widehat{\mathbf{N}}}
		\global\long\def\hatmbo{\widehat{\mathbf{o}}}
		\global\long\def\hatmbO{\widehat{\mathbf{O}}}
		\global\long\def\hatmbp{\widehat{\mathbf{p}}}
		\global\long\def\hatmbP{\widehat{\mathbf{P}}}
		\global\long\def\hatmbq{\widehat{\mathbf{q}}}
		\global\long\def\hatmbQ{\widehat{\mathbf{Q}}}
		\global\long\def\hatmbr{\widehat{\mathbf{r}}}
		\global\long\def\hatmbR{\widehat{\mathbf{R}}}
		\global\long\def\hatmbs{\widehat{\mathbf{s}}}
		\global\long\def\hatmbS{\widehat{\mathbf{S}}}
		\global\long\def\hatmbt{\widehat{\mathbf{t}}}
		\global\long\def\hatmbT{\widehat{\mathbf{T}}}
		\global\long\def\hatmbu{\widehat{\mathbf{u}}}
		\global\long\def\hatmbU{\widehat{\mathbf{U}}}
		\global\long\def\hatmbv{\widehat{\mathbf{v}}}
		\global\long\def\hatmbV{\widehat{\mathbf{V}}}
		\global\long\def\hatmbw{\widehat{\mathbf{w}}}
		\global\long\def\hatmbW{\widehat{\mathbf{W}}}
		\global\long\def\hatmbx{\widehat{\mathbf{x}}}
		\global\long\def\hatmbX{\widehat{\mathbf{X}}}
		\global\long\def\hatmby{\widehat{\mathbf{y}}}
		\global\long\def\hatmbY{\widehat{\mathbf{Y}}}
		\global\long\def\hatmbz{\widehat{\mathbf{z}}}
		\global\long\def\hatmbZ{\widehat{\mathbf{Z}}}

		\global\long\def\bolalpha{\boldsymbol{\alpha}}
		\global\long\def\bolbeta{\boldsymbol{\beta}}
		\global\long\def\bolgamma{\boldsymbol{\gamma}}
		\global\long\def\boldelta{\boldsymbol{\delta}}
		\global\long\def\bolepsilon{\boldsymbol{\epsilon}}
		\global\long\def\bolvarepsilon{\boldsymbol{\varepsilon}}
		\global\long\def\bolzeta{\boldsymbol{\zeta}}
		\global\long\def\boleta{\boldsymbol{\eta}}
		\global\long\def\boltheta{\boldsymbol{\theta}}
		\global\long\def\bolkappa{\boldsymbol{\kappa}}
		\global\long\def\bollambda{\boldsymbol{\lambda}}
		\global\long\def\bolmu{\boldsymbol{\mu}}
		\global\long\def\bolnu{\boldsymbol{\nu}}
		\global\long\def\bolxi{\boldsymbol{\xi}}
		\global\long\def\bolpi{\boldsymbol{\pi}}
		\global\long\def\bolrho{\boldsymbol{\rho}}
		\global\long\def\bolsigma{\boldsymbol{\sigma}}
		\global\long\def\boltau{\boldsymbol{\tau}}
		\global\long\def\bolphi{\boldsymbol{\phi}}
		\global\long\def\bolchi{\boldsymbol{\chi}}
		\global\long\def\bolpsi{\boldsymbol{\psi}}
		\global\long\def\bolomega{\boldsymbol{\omega}}
		\global\long\def\bolGamma{\boldsymbol{\Gamma}}
		\global\long\def\bolDelta{\boldsymbol{\Delta}}
		\global\long\def\bolTheta{\boldsymbol{\Theta}}
		\global\long\def\bolLambda{\boldsymbol{\Lambda}}
		\global\long\def\bolPi{\boldsymbol{\Pi}}
		\global\long\def\bolSigma{\boldsymbol{\Sigma}}
		\global\long\def\bolPhi{\boldsymbol{\Phi}}
		\global\long\def\bolPsi{\boldsymbol{\Psi}}
		\global\long\def\bolOmega{\boldsymbol{\Omega}}
				\global\long\def\bolXi{\boldsymbol{\Xi}}

		\global\long\def\hatbolalpha{\widehat{\boldsymbol{\alpha}}}
		\global\long\def\hatbolbeta{\widehat{\boldsymbol{\beta}}}
		\global\long\def\hatbolgamma{\widehat{\boldsymbol{\gamma}}}
		\global\long\def\hatboldelta{\widehat{\boldsymbol{\delta}}}
		\global\long\def\hatbolepsilon{\widehat{\boldsymbol{\epsilon}}}
		\global\long\def\hatbolzeta{\widehat{\boldsymbol{\zeta}}}
		\global\long\def\hatboleta{\widehat{\boldsymbol{\eta}}}
		\global\long\def\hatboltheta{\widehat{\boldsymbol{\theta}}}
		\global\long\def\hatbolkappa{\widehat{\boldsymbol{\kappa}}}
		\global\long\def\hatbollambda{\widehat{\boldsymbol{\lambda}}}
		\global\long\def\hatbolmu{\widehat{\boldsymbol{\mu}}}
		\global\long\def\hatbolnu{\widehat{\boldsymbol{\nu}}}
		\global\long\def\hatbolxi{\widehat{\boldsymbol{\xi}}}
		\global\long\def\hatbolpi{\widehat{\boldsymbol{\pi}}}
		\global\long\def\hatbolrho{\widehat{\boldsymbol{\rho}}}
		\global\long\def\hatbolsigma{\widehat{\boldsymbol{\sigma}}}
		\global\long\def\hatboltau{\widehat{\boldsymbol{\tau}}}
		\global\long\def\hatbolphi{\widehat{\boldsymbol{\phi}}}
		\global\long\def\hatbolchi{\widehat{\boldsymbol{\chi}}}
		\global\long\def\hatbolpsi{\widehat{\boldsymbol{\psi}}}
		\global\long\def\hatbolomega{\widehat{\boldsymbol{\omega}}}
		\global\long\def\hatbolGamma{\widehat{\boldsymbol{\Gamma}}}
		\global\long\def\hatbolDelta{\widehat{\boldsymbol{\Delta}}}
		\global\long\def\hatbolTheta{\widehat{\boldsymbol{\Theta}}}
		\global\long\def\hatbolLambda{\widehat{\boldsymbol{\Lambda}}}
		\global\long\def\hatbolPi{\widehat{\boldsymbol{\Pi}}}
		\global\long\def\hatbolSigma{\widehat{\boldsymbol{\Sigma}}}
		\global\long\def\hatbolPhi{\widehat{\boldsymbol{\Phi}}}
		\global\long\def\hatbolPsi{\widehat{\boldsymbol{\Psi}}}
		\global\long\def\hatbolOmega{\widehat{\boldsymbol{\Omega}}}

		\global\long\def\barbolmu{\overline{\bolmu}}
		\global\long\def\barmbX{\overline{\mbX}}

		\global\long\def\mbbR{\mathbb{R}}
		\global\long\def\mbbP{\mathbb{P}}
		\global\long\def\mbbQ{\mathbb{Q}}
		\global\long\def\mbbS{\mathbb{S}}
		\global\long\def\mbbH{\mathbb{H}}
		\global\long\def\mbbX{\mathbb{X}}
		\global\long\def\mbbY{\mathbb{Y}}
		\global\long\def\mbbZ{\mathbb{Z}}
		\global\long\def\spc{\mathcal{S}}

		\global\long\def\calA{\mathcal{A}}
		\global\long\def\calB{\mathcal{B}}
		\global\long\def\calC{\mathcal{C}}
		\global\long\def\calD{\mathcal{D}}
		\global\long\def\calE{\mathcal{E}}
		\global\long\def\calF{\mathcal{F}}
		\global\long\def\calG{\mathcal{G}}
		\global\long\def\calH{\mathcal{H}}
		\global\long\def\calI{\mathcal{I}}
		\global\long\def\calJ{\mathcal{J}}
		\global\long\def\calK{\mathcal{K}}
		\global\long\def\calL{\mathcal{L}}
		\global\long\def\calM{\mathcal{M}}
		\global\long\def\calN{\mathcal{N}}
		\global\long\def\calO{\mathcal{O}}
		\global\long\def\calP{\mathcal{P}}
		\global\long\def\calQ{\mathcal{Q}}
		\global\long\def\calR{\mathcal{R}}
		\global\long\def\calS{\mathcal{S}}
		\global\long\def\calT{\mathcal{T}}
		\global\long\def\calU{\mathcal{U}}
		\global\long\def\calV{\mathcal{V}}
		\global\long\def\calW{\mathcal{W}}

		\global\long\def\mbell{\boldsymbol{\ell}}
		\global\long\def\bolell{\boldsymbol{\ell}}
		\global\long\def\mbzero{\mathbf{0}}

		\global\long\def\bolPhio{\boldsymbol{\Phi}_{0}}
		\global\long\def\bolOmegao{\boldsymbol{\Omega}_{0}}

		\global\long\def\bolSigmaX{\bolSigma_{\mbX}}
		\global\long\def\bolSigmaY{\bolSigma_{\mbY}}
		\global\long\def\bolSigmaXY{\boldsymbol{\Sigma}_{\mbX\mbY}}
		\global\long\def\mbSX{\mathbf{S}_{\mbX}}
		\global\long\def\mbSY{\mathbf{S}_{\mbY}}
		\global\long\def\mbSXY{\mathbf{S}_{\mbX\mbY}}
		\global\long\def\mbSYX{\mathbf{S}_{\mbY\mbX}}
		\global\long\def\mbRYX{\mathbf{S}_{\mbY|\mbX}}
		\global\long\def\mbRXY{\mathbf{S}_{\mbX|\mbY}}
		\global\long\def\mbSc{\mbS_{\mbC}}
		\global\long\def\mbSd{\mbS_{\mbD}}

		\global\long\def\sumn{\sum_{i=1}^{n}}

		\global\long\def\E{\mathrm{E}}
		\global\long\def\F{\mathrm{F}}
		\global\long\def\J{\mathrm{J}}
		\global\long\def\H{\mathrm{H}}
		\global\long\def\G{\mathrm{G}}
		\global\long\def\Cov{\mathrm{cov}}
		\global\long\def\cov{\mathrm{cov}}
		\global\long\def\Corr{\mathrm{corr}}
		\global\long\def\Var{\mathrm{var}}
		\global\long\def\dimension{\mathrm{dim}}
		\global\long\def\spn{\mathrm{span}}
		\global\long\def\vech{\mathrm{vech}}
		\global\long\def\vecc{\mathrm{vec}}
		\global\long\def\Prob{\mathrm{Pr}}
		\global\long\def\Env{\mathrm{env}}
		\global\long\def\tr{\mathrm{tr}}
		\global\long\def\dg{\mathrm{diag}}
		\global\long\def\asyVar{\mathrm{avar}}
		\global\long\def\MSE{\mathrm{MSE}}
		\global\long\def\OLS{\mathrm{OLS}}
		\global\long\def\mm{\mathrm{M}}
		\global\long\def\im{\mathrm{IM}}
		\global\long\def\tvec{\text{vec}}
		\global\long\def\tvech{\text{vech}}
		\global\long\def\bma{\bm{a}}
		\global\long\def\bmA{\bm{A}}
		\global\long\def\bmb{\bm{b}}
		\global\long\def\bmB{\bm{B}}
		\global\long\def\bmc{\bm{c}}
		\global\long\def\bmC{\bm{C}}
		\global\long\def\bmd{\bm{d}}
		\global\long\def\bmD{\bm{D}}
		\global\long\def\bme{\bm{e}}
		\global\long\def\bmE{\bm{E}}
		\global\long\def\bmf{\bm{f}}
		\global\long\def\bmF{\bm{F}}
		\global\long\def\bmg{\bm{g}}
		\global\long\def\bmG{\bm{G}}
		\global\long\def\bmh{\bm{h}}
		\global\long\def\bmH{\bm{H}}
		\global\long\def\bmi{\bm{i}}
		\global\long\def\bmI{\bm{I}}
		\global\long\def\bmj{\bm{j}}
		\global\long\def\bmJ{\bm{J}}
		\global\long\def\bmk{\bm{k}}
		\global\long\def\bmK{\bm{K}}
		\global\long\def\bml{\bm{l}}
		\global\long\def\bmL{\bm{L}}
		\global\long\def\bmm{\bm{m}}
		\global\long\def\bmM{\bm{M}}
		\global\long\def\bmn{\bm{n}}
		\global\long\def\bmN{\bm{N}}
		\global\long\def\bmo{\bm{o}}
		\global\long\def\bmO{\bm{O}}
		\global\long\def\bmp{\bm{p}}
		\global\long\def\bmP{\bm{P}}
		\global\long\def\bmq{\bm{q}}
		\global\long\def\bmQ{\bm{Q}}
		\global\long\def\bmr{\bm{r}}
		\global\long\def\bmR{\bm{R}}
		\global\long\def\bms{\bm{s}}
		\global\long\def\bmS{\bm{S}}
		\global\long\def\bmt{\bm{t}}
		\global\long\def\bmT{\bm{T}}
		\global\long\def\bmu{\bm{u}}
		\global\long\def\bmU{\bm{U}}
		\global\long\def\bmv{\bm{v}}
		\global\long\def\bmV{\bm{V}}
		\global\long\def\bmw{\bm{w}}
		\global\long\def\bmW{\bm{W}}
		\global\long\def\bmx{\bm{x}}
		\global\long\def\bmX{\bm{X}}
		\global\long\def\bmy{\bm{y}}
		\global\long\def\bmY{\bm{Y}}
		\global\long\def\bmz{\bm{z}}
		\global\long\def\bmZ{\bm{Z}}
		\global\long\def\CS{\calS_{Y\mid\mbX}}
		\global\long\def\covenv{\calE_{\bolDelta_{0}}(\bolbeta)}
		\global\long\def\adj{\text{adj}}